\documentclass{an}
\usepackage{times}
\usepackage{graphicx}

\begin{document}
%
%
\title{Structure and evolution of pulsating hot subdwarfs} 
\author{Steven D. Kawaler} 
\institute{Department of Physics and Astronomy, Iowa State University, Ames, IA  50011  USA} 

\abstract{Hot subdwarfs are evolved low--mass stars that have survived core helium ignition and are now in (or recently
finished with) the core helium burning stage. At the hot end of the Horizontal Branch (HB), many
of these stars are multiperiodic pulsators. These pulsations have revealed details of their
global and internal structure, and provide important constraints on the origin of hot HB stars. While many features of their structure deduced from seismic
fits have confirmed what we expected from evolutionary considerations, there have been
some surprises as well.
} 
\maketitle

\section{Introduction}

The fundamental picture of low-mass stellar evolution includes the still--enigmatic  transition from the red giant phase to the helium core burning phase.  Stars that begin life with masses less than about 2.3~M$_{\odot}$ enter this phase through ignition of helium while the core is degenerate - the core helium flash (first studied by  Mestel 1952 and Schwarzschild \& H\"arm 1962).  Core helium-burning models that have survived the flash populate the Horizontal Branch (HB), as as noted first by Hoyle \& Schwarzschild (1955). 

Historically, computation of evolution through the core helium flash has been a difficult problem.  To get around the core helium flash problem, HB models are generally constructed as ``Zero Age Horizontal Branch'' (ZAHB) models using a pure helium core surrounded by a hydrogen-rich envelope.  To account for the distribution of core helium-burning stars in the H--R diagram, the stars must lose a significant amount of mass at (or before) the onset of the flash, as was first demonstrated numerically in the last 1960s (Faulkner \& Iben 1966, Rood 1970, Iben \& Rood 1970).  Families of ZAHB models share the same core (the mass determined by the size of the hydrogen--exhausted core at the onset of the helium flash, with the amount of remnant envelope is treated as a free parameter.  The thinner the hydrogen surface layer, the higher the effective temperature.

This approach to modeling HB stars has remain largely unchanged over the past 40 years.  This way of making models of stars avoids the helium core flash questions, yet provides satisfactory (though not complete) models of HB stars in globular clusters and other old stellar populations.  However, we learn little about the conditions of the core helium flash or about the mass loss event(s) that reduce the mass of the envelope when modeling stars in this traditional way.

What has changed over the past 10-15 years is the increased interest in the origin and evolution of the hottest horizontal branch (EHB) stars, such as the hot subdwarf B (sdB) stars.   To account for the bluest EHB stars, the remnant hydrogen layer must be thinner than about 0.5\% of the star's mass.  Stars with envelopes that thin (or thinner) do not support hydrogen shell burning - the entire luminosity of the star is generated by helium core burning. Figuring out how a star can leave itself with such a thin hydrogen layer has perplexed astronomers for decades. As far back as 1976, Mengel et al. (1976) proposed binary mass exchange as one possible mechanism; other proposals for their origin include variations of mass loss 
efficiency on the RGB (D'Cruz et al. 1996), 
and even mass stripping by planetary companions (Soker \& Harpaz 2000).
Growing observational evidence now suggests that binarity plays a role in the formation of the subdwarf B (sdB) stars (Maxted et al. 2001, Heber 2009).  Theoretical support for this view comes from a series of simulations of binary evolution (Sandquist et al. 2000, Han et al. 2002, 2003).

Over this same time, discovery that some sdB stars undergo nonradial pulsation has brought the tools of asteroseismology to the study of these stars.  Two broad classes are known -  short--period $p$-mode pulsators, and longer--period $g$-mode pulsators. and the frequencies provide a probe of their global properties and internal properties.  These analyses have concentrated on as well as structural parameters such as the thickness of the surface hydrogen-rich layer via matching the theoretical $p$-mode frequencies of models to the observations.  There are several excellent recent reviews of this exciting field -- see, for example, Charpinet et al. (2008a) and {\O}stensen (2010).  For a general review of observations HB stars, see the excellent review by Catelan (2009), and for a recent account of all things hot subddwarf, see Heber (2009).

In this brief review, I recount the basic theoretical aspects of sdB interior evolution, and focus on how asteroseismology addresses questions about the origin of sdB stars, and more generally, the physical processes that operate in the cores of HB stars.  Seismic analysis of $p$-modes has revealed the envelope structure of many sdB stars.  But some of these $p$-mode pulsators, along with a growing number of longer period $g$-mode pulsators, allow us to pursue the goal of probing the cores of sdB stars.  In that way we can ``look'' at the region of the star that was directly involved in the core helium flash. The nonradial pulsation spectrum in this subgroup of stars can help us test these models and see the physics in action.

\section{Models of sdB evolution}

Ordinary HB stars (those that have envelopes thick enough to support hydrogen--shell burning) can arrive on the ZAHB through single star evolution in a relatively simple way, such as through steady mass loss on the RGB, or a single episode of modest mass loss at he core helium flash.  Arrival on the EHB with very thin envelopes is more problematic.  Fine-tuning of the mass loss mechanism is needed to produce such a thin hydrogen--rich envelope.  However, single--star scenarios that result in EHB stars produce structures that are entirely analogous to those with thicker envelopes.

\subsection{Standard model(s)}

\begin{figure} 
\includegraphics[width=8cm]{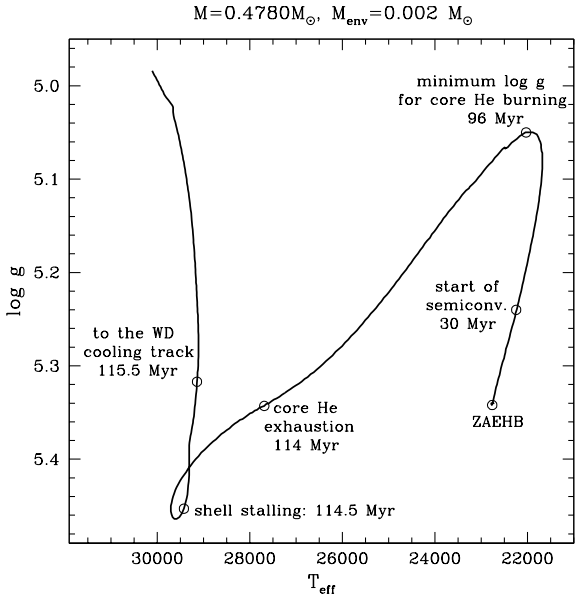} 
\caption{Anatomy of a representative sdB evolutionary track for a model with a mass of 0.478M$_{\odot}$.  Evolutionary stages and the time following the Zero Age EHB are indicated. At the start of semiconvection, the core helium abundance has been reduced to 0.66.  When the model reaches the minimum $\log g$, that abundance has dropped to 0.12.  At that point the model has a fully convective core out to a mass fraction of 0.28~M$_*$ and the top of the semiconvective region is at a mass fraction of 0.5~M$_*$.} 
\label{fig01} 
\end{figure}

For standard sdB models -- with hydrogen-rich envelopes that are too thin to support a hydrogen burning shell -- the evolutionary tracks have a distinctive shape.  A representative example is shown in Fig.\ref{fig01}, constructed as described in Kawaler \& Hostler (2005).  Increasing (decreasing) the thickness of the surface hydrogen layer translates the curve towards lower (higher) temperatures and gravities.

Upon settling on the ZAHB, the core has commenced burning helium via the 3$\alpha$ process.  The basic time scale for core helium burning in these stars is about $10^8$ years.   This highly temperature--sensitive reaction results in vigorous core convection that generates a steep composition gradient at the edge of the convective core.  One consequence of the steep composition gradient is that the convection boundary is unstable to small amounts of outward mixing.  Thus, by 30~Myr after core helium ignition, the core is likely to overshoot, resulting in extra mixing.  If it occurs gently, this extra mixing resembles diffusive mixing, where the composition gradient is ``softened'' to ensure neutral stability against convection -- a process often called {\it semiconvection}.  The precise mechanism by which this extra mixing occurs is not fully understood, even if it can be computed conveniently through mixing to convective neutrality.

\begin{figure} 
\includegraphics[width=8.4cm]{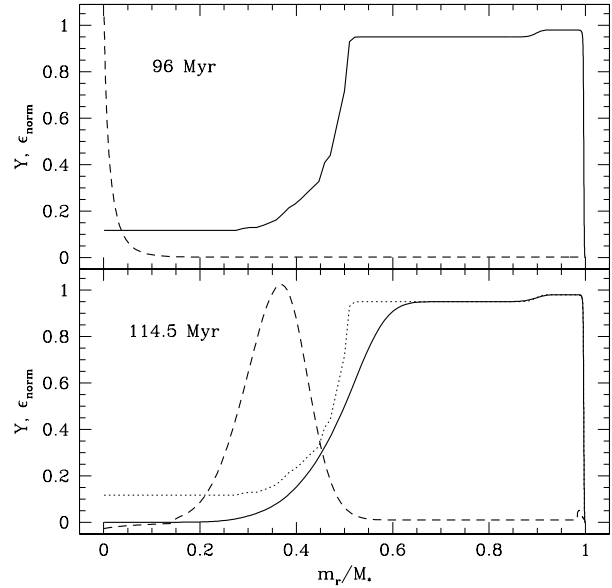} 
\caption{Internal structure of two sdB models.  {\bf Top:} helium abundance (solid line) and normalized energy generation rate per gram (dashed line) for the 96~Myr model that is nearing core helium exhaustion.  {\bf Bottom:} the same quantities during the shell narrowing phase - the energy generation rate peaks in the narrowing helium-burning region surrounding the C/O core.  The dotted line in the bottom panel shows the composition profile of the upper model for comparison.} 
\label{fig02} 
\end{figure}

Through most of helium core burning, the effective temperature of the model stays roughly constant, but the star grows in radius and therefore moves upwards in the H--R diagram (see Fig.~\ref{fig01}).  As the core becomes significantly depleted in helium, the evolutionary track turns around, and the model moves more quickly to higher temperature and gravity.  The luminosity and helium abundance profile at the turning point are shown in the upper panel of Fig.~\ref{fig02}.  At this point, the fully convective core reaches out to a mass fraction of 0.3M$_*$ (or 0.14M$_{\odot}$), while the semiconvective zone reaches the midpoint in mass (0.5M$_*$ or 0.24M$_{\odot}$).  

Following helium exhaustion in the core, a thick helium--burning shell develops, working its way quickly out from the center as it consumes the surrounding, helium-depleted material.  As the shell encounters the helium-rich material, it slows, and the luminosity (and radius) begin to increase (the shell--stalling phase in Fig.~\ref{fig01}) and the model makes its way, eventually, to the white dwarf cooling track.

\subsection{Evolution from binary progenitors}

As mentioned above, a leading candidate for the production mechanism for sdB stars involves common--envelope evolution, leading to mass transfer and stripping of the bulk of the hydrogen--rich envelope.  Stars produced via this mechanism can arrive in the sdB region of the H--R diagram with total masses and envelope masses that are significantly different than for single--star production scenarios (Han et al. 2002, 2003, Hu et al. 2008).

Hu et al. (2008) explore evolution to and beyond the ZAEHB using higher-mass initial models that have had mass stripped while on the RGB (as a consequence of binary evolution).  Their initial mass is sufficiently high that they ignite helium non--degenerately, avoiding the helium core flash but still evolving to the hot horizontal branch.  Their subsequent evolution includes an active hydrogen--burning shell, and their evolutionary path is quite different than the standard model described above (see Hu et al. 2008 for details).  

\subsection{Diffusion}

Because of their relatively high surface gravities and radiative envelopes, sdB stars are subject to diffusive properties that can significantly alter their surface abundances and subsurface composition profiles.  Observational evidence includes  significant helium depletion in EHB stars by a factor of 10 to 100 (Heber et al. 1986) which likely results from gravitational settling.  Theoretical calculations can explain the general trends suspected for diffusion (depletion of helium and some heavier elements) illustrating the importance of diffusion on the evolutionary paths of these stars for both standard models and those with binary progenitors (i.e. Hu et al. 2010).

In addition to gravitational settling, another important process is radiative levitation of metallic species (Fontaine \& Chayer 1997, Moehler et al. 2000, Unglaub \& Bues 2001), which can account for peculiar distribution of these elements in the spectra of these stars (Geier et al. 2010).  Inclusion of even very small amounts of stellar wind can also alter the surface abundances (Fontaine \& Chayer 1997, Unglaub \& Bues 2001, Chayer et al. 2004, Unglaub 2008) and in fact may be necessary to understanding the detailed abundance patterns.  Models that include all of these diffusive processes also predict that iron-peak elements should also accumulate below the surface of these stars which turns out to be crucial to explaining the driving mechanism for the pulsating sdB stars (Charpinet et al. 1997).

\section{Seismology of sdB stars - envelopes}

Our desire to understand the origin of sdB stars requires more information than can be provided by traditional spectroscopic observation. Such observations yield precise positions in the H--R diagram which can be compared, both individually and as an ensemble, to evolutionary tracks.  Time--resolved spectroscopy and photometry can also reveal the presence (and properties) of binary companions as well. But it is hard to construct a detailed make-or-break experiment that could be decisive in confirming a production scenario that is based on this approach.

However, as shown by Hu et al. (2008), the internal structure of an sdB star can reveal its progenitor by its influence on the observed $p$-mode spectrum.  We can, for example, use asteroseismic probes will be able to measure the position below the surface of the H/He transition layer for comparison to models with different origins.

In addition, these probes also allow us to test models of diffusion, radiative levitation, and winds in sdB stars.  Combined with spectroscopic measurements of elemental abundances, this powerful suite of observational tools can give us an empirical picture of the physics operating within the envelopes of sdB stars.

\subsection{Hydrogen layer thickness - adiabatic asteroseismology}

The short-period sdB stars are pulsating in $p$-modes, which sample the surface layers much more sensitively than the core.  Therefore, comparison between observed pulsation frequencies and models (constrained by spectroscopically determined values of $T_{\rm eff}$ and $\log g$) can reveal details about the surface layers (notably, $M_H$) as well as the global properties (mass, radius, and luminosity) of the stars.

The most exhaustive efforts at seismic probing of the short period sdB pulsators is the work over the past decade by Charpinet, Fontaine, Brassard, and their collaborators.  They have analyzed seismic data on more than a dozen of the short--period sdB stars, and determined estimates of the total mass and surface hydrogen layer thicknesses for most of those stars. A good recent summary of the many results from their analysis can be found in Fontaine et al. (2008); see {\O}stensen (2010) for an updated summary plot showing the distribution of resulting core/envelope masses.

The determined range of masses is wider than would be expected for single star evolution, suggesting that the binary formation channel is indeed an important contributor to the observed sdB population. While further work needs to be done to estimate the relative contribution of the binary channel(s) this result demonstrates the potential of asteroseismology to address this fundamental question about the origin of sdB stars.  An additional important conclusion is that the envelope thickness determinations are, in all cases, as small as the standard model suggests - indicating that, as expected, the surface hydrogen layer in these stars is too thin to support a hydrogen burning shell.

\subsection{Diffusion and ``nonadiabatic'' asteroseismology}

Charpinet et al. (1997) were the first to show the importance of subsurface iron in driving pulsations of sdB stars (see also Fontaine et al. 2006, Charpinet et al. 2008a, and references therein). Without radiative levitation, the iron abundance is too low for the iron-peak opacity to drive the oscillations.  Thus the very existence of pulsations in sdB stars is evidence that radiative levitation of heavy elements is a real process in these stars.  The fact that iron-peak elements are enhanced below the surface does not mean that the surface abundances are necessarily enhanced, as evidenced by the observations by O'Toole \& Heber (2006).

The models are robust and predictive: unstable pulsations require levitated iron near (but below) the surface at the location in the H--R diagram where we find pulsators.  This permits the blue (and red) edges of the instability region to constrain the internal iron distribution in that models that are unstable should match (in $T_{\rm eff}$ and $\log g$ the observed instability region.  This form of ``nonadiabatic asteroseismology'' is explored by Charpinet et al. (2008a, 2009), who show that models with radiative levitation do indeed match the observed instability region for the short-period sdB pulsators.

\section{Core properties of sdB stars via seismology}

While much of the activity in studying the seismology of sdB stars has been concentrated on the $p$-mode pulsators and diagnostics of global and envelope properties, the long--period pulsators hold promise for studying the cores of sdB stars.  This has been a relatively limited area because of the difficulties of observing and resolving dense pulsation spectra with periods of thousands of seconds (see, for example, the heroic effort by Randall et al. 2006), but {\it Kepler} observations of these stars will provide ample data for $g$-mode probes of their cores.  Stay tuned!

In the interim, those sdB pulsators with periods in the 300s to 600s range (such as Feige 48, PG 1605+072, V391 Peg, and Balloon 090100001) do give us some hope of testing models of their cores.  Measurement of rates of period change in these stars can constrain their evolutionary status and therefore provide information on their core configurations.  Also, for evolved sdB stars (beyond helium core exhaustion) many of the observed longer-period modes are of mixed character - that is, they behave as $g$-modes in the core but $p$-modes in the envelope.  Within their observed period range, there are a mixture of modes that sample very different depths.  If we can identify those modes, they can provide seismic probes of the core.

\subsection{Rate of period change - secular evolution}

One of the most exciting side--results of sdB seismology was the discovery by Silvotti et al. (2007 - hereafter S07) of reflex orbital motion induced by a planetary companion.  While the planet discovery is truly important, the measurement of the rate of period change ($\dot{P}$ or $dP/dt$) for two independent modes in the star is also notable as the first such measurement in an sdB star.  The values for the two modes are shown in Table~\ref{tab01}.

In the short--period sdB stars, since the pulsation frequencies scale inversely with the square root of the density $\dot{P}$ is determined by the rate of contraction (or expansion):  
\begin{equation}
P \propto \frac{1}{\sqrt{G \rho}} \propto R^{3/2} (GM)^{-1/2} \\
{\rm so}\\
\frac{\dot{P}}{P} \approx \frac{3}{2} \frac{\dot{R}}{R} \\.
\end{equation}
In addition to this general scaling, the values of $\dot{P}$ also depend on the eigenfunctions. The shape of the eigenfunctions determine what parts of the star are most important in determining the pulsation frequencies and the rates at which they change.  For V391 Peg, the values of $\dot{P}$ differ by 2$\sigma$.  

For V391~Peg, the time scale for period change ($P/\dot{P}$) is roughly +6.5$\times 10^6$y as measured by S07.  This corresponds to a time scale for radius {\it increase} of approximately $10^7$y, which is much faster than the evolutionary time scale for sdB stars that are burning helium in the core.  Furthermore, it is inconsistent with any evolutionary phase where the stellar radius is decreasing with time.

\begin{figure} 
\includegraphics[width=8.4cm]{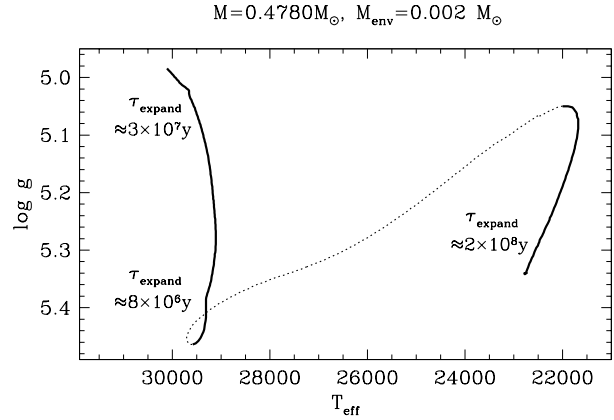} 
\caption{Evolutionary track for the same sdB model sequence as in Fig.\ref{fig01}.  The bold portions of the track denote times when the radius of the model is increasing. The time scale for expansion, $R_*/\dot{R}$, is indicated.  During core helium burning, the radius increases on a time scale that is an order of magnitude slower than the expansion rate following the establishment of the helium-burning shell.} 
\label{fig03} 
\end{figure}

In Fig.~\ref{fig03} we show the same evolutionary track as in Fig.~\ref{fig01}, with highlighted portions indicating evolution with increasing radius, along with the time scale for radius increase.  To match the observed values of $\dot{P}$ of V391~Peg, the models suggest that this star is well beyond the core helium burning phase, and is evolving towards the white dwarf cooling phase with a helium burning shell surrounding the helium--exhausted core.  

It is interesting to note that models can be found that match the two main periods in V391~Peg, as well as the observed $T_{\rm eff}$ and $\log g$, but which are in very different evolutionary stages.  Table~\ref{tab01} shows two representative models - one burning helium in a shell (similar to the 114.5~Myr model in Figs.~\ref{fig01} and \ref{fig02}) and one still burning helium in the core (similar to the 96~Myr model but with a thinner hydrogen shell).  However, to match the observed values of $\dot{P}$, the shell--burning model is very close while the core--burning model is changing too slowly (by a factor of 100).  Thus measured rates of period change can place important constraints on the evolutionary status of the sdB pulsators.  

\begin{table}
\caption{Periods, rates of period change, and model comparisons for V391~Peg.\label{tab01}}
\begin{tabular}{lccc}
\hline
Quantity                &  Observed   & Shell-burnning  & Core-burning \\
                        & (S07) & 114.8 Myr & 88.3 Myr\\
\hline
$T_{\rm eff}$              & $29300\pm500$ &  29219    &  28395    \\
$\log g$                   & $5.4\pm0.1$    &  5.358     &   5.507    \\
$M_*$                   &               & 0.4780M$_{\odot}$ &  0.4902M$_{\odot}$\\
$M_{\rm H}$             &               & 0.0020M$_{\odot}$ &  0.0002M$_{\odot}$\\
 & & & \\
$P_1$                     & 349.54~s        &  348.69~s    &  349.51~s    \\
$\dot{P}$ ($10^{-12}$) & +$1.46\pm0.07$  & +1.55      & +0.028     \\
$P/\dot{P}$          & 7.6$\times 10^6$y & 7.2$\times 10^6$y &4.0$\times 10^8y$    \\
$(l,n)$                       &            & $(1,-3)$  & $(1,1)$ \\
                          & & & \\
$P_2$                     & 354.10~s         & 352.96~s     &  353.14~s    \\
$\dot{P}$ ($10^{-12}$) & +$2.05\pm0.26$  & +2.67      &  +0.029    \\
$P/\dot{P}$       & 5.5$\times10^6$y  &  4.2$\times10^6$y  &  3.9$\times10^8$y     \\
$(l,n)$                   &                    & (0,1) & (0,0) \\
\hline
\end{tabular}
\end{table}

While this sketch of the way that $\dot{P}$ can help constrain modes looks promising, it is important to also note that the measure rates of period change will always be upper limits to the evolutionary rates of the stars.  All secular processes occurring within the stars operate on the longest available time scales, so any process that might influence the phase or frequency of a mode will push the resulting determination of $\dot{P}$ to faster changes.  Nonlinear mode coupling, stochastic phase variations or other ``weather--like'' phenomena could make it appear that the star is evolving more quickly.  Therefore, the conclusion that V319 Peg is highly evolved requires more precise modeling and (hopefully) observations of $\dot{P}$ for additional modes.

\subsection{Mixed-character modes}

While the short-period sdB variables are generally considered to be $p$-mode pulsators, with the oscillations being insensitive to core properties, the lower-gravity members of this class generally show longer-period pulsations (i.e. in excess of 300~s).  For these stars, the models indicate that some of the modes are mixed--character, as has been shown by Kawaler (1999) and, more extensively, by Charpinet et al. (2000) and Hu et al. (2009).  Some modes have oscillation eigenfunctions have nodes close to the center and behave like $g-$modes there, but also have a node or nodes near the surface, where the eigenfunction is more $p$-mode-like.  That means that some the oscillations in these stars can allow us to probe the core.

\begin{figure} 
\includegraphics[width=8cm]{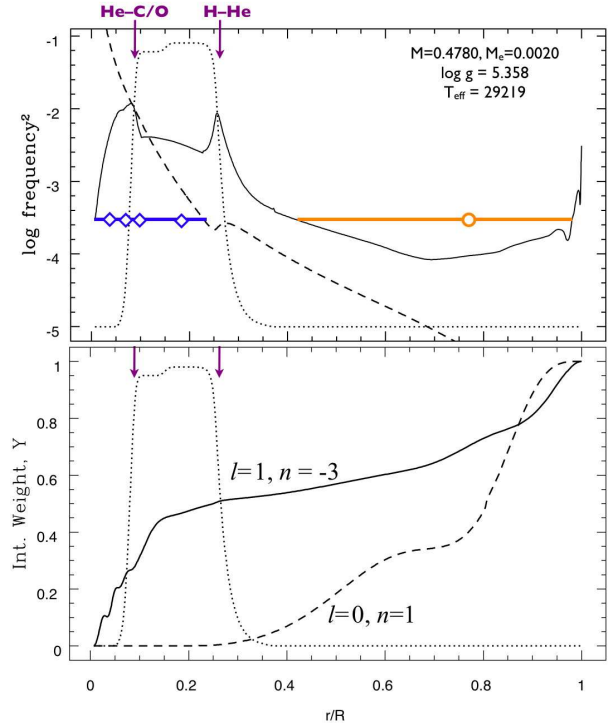} 
\caption{Properties of mixed-character modes in an evolved sdB model that has exhausted helium in its core.  {\bf Top:} the propagation diagram, with the Brunt-V\"ais\"al\"a frequency (solid line) and Lamb frequency (dashed line) as a function of fractional radius.  The horizontal line lies at the frequency corresponding to the $l$=1, $n$=-3 $g$-mode; diamonds denote nodes in the radial eigenfunction that are $g$-mode-like, and the circle shows the envelope node in the $p$-mode propagation region. {\bf Bottom:} the integrated adiabatic period weight functions.  For the $l$=0, $n$=1 radial mode, the node lies at $r/R=0.77$. The period of pulsation is 348.7s for the $l$=1 mode, and 353.0s for the $l$=0 mode. In both panels, the dotted line shows the helium composition (with vertical scale from 0 to 1). } 
\label{fig04} 
\end{figure}

Fig.~\ref{fig04} shows some internal properties of a model of such a star - the model that provides a match to the period and $\dot{P}$ of V391~Peg listed in Table~\ref{tab01}.  In the top panel, the Brunt--V\"ais\"al\"a frequency shows two bumps at the position of the H/He transition and at the edge of the He--exhausted core.  For the $l$=1, $n$=-3 mode, the inner composition transition acts as a reflecting boundary, partially trapping a mode in the interior, and enhancing the period weight function significantly in that region. Even though that mode has a frequency that is very similar to the frequency of the first overtone radial mode, it is in essence a $g$-mode.

\begin{figure} 
\includegraphics[width=5.8cm,angle=270,clip]{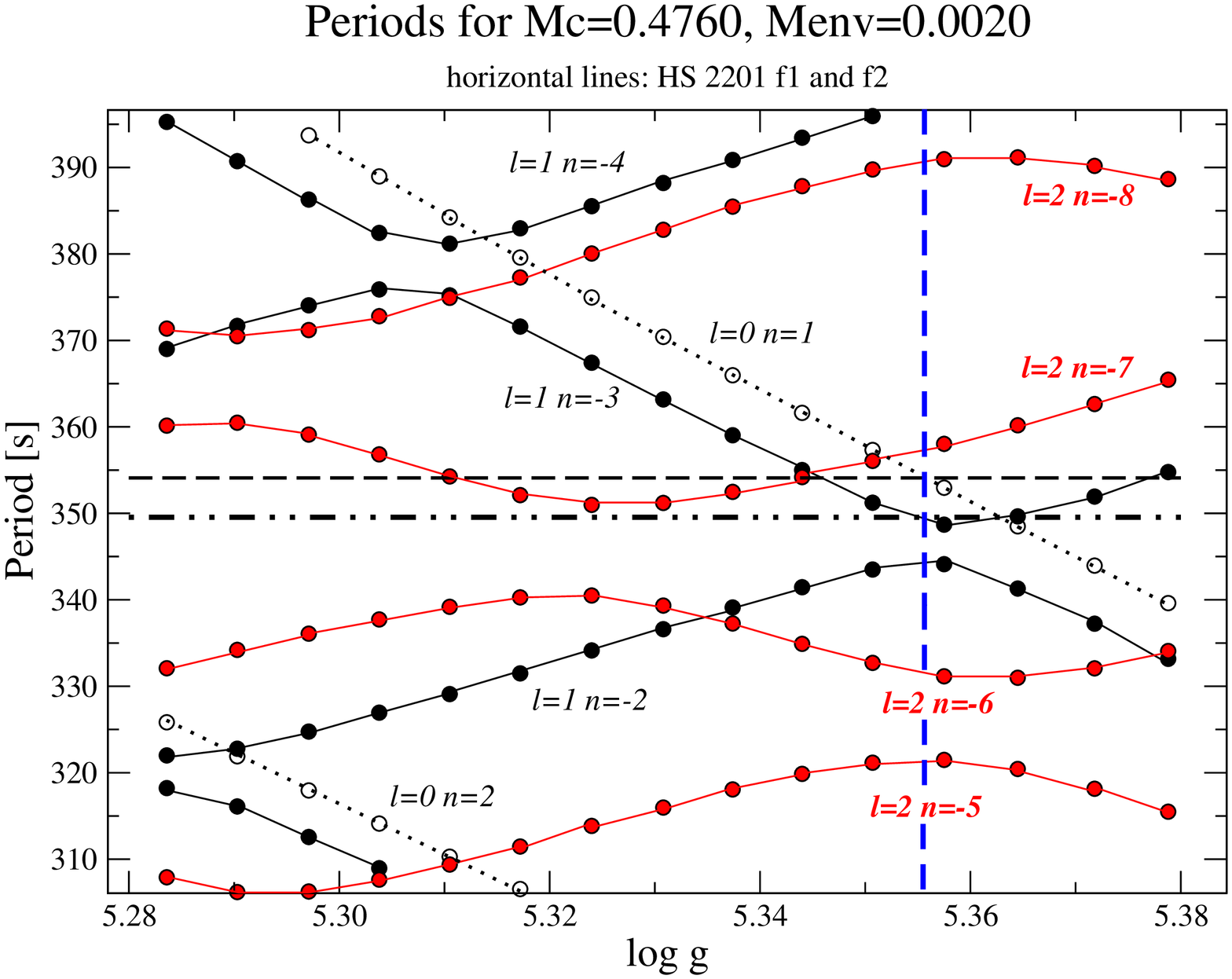} 
\caption{Avoided crossings in an evolving, helium-shell burning sdB model. Period is shown vs. $\log g$, and the model evolves from right to left.  Horizontal lines indicate periods of the two main modes in V391 Peg.  Each vertical string of dots represents periods in a model at a given $\log g$; lines connect modes with same values of $n$ and $l$.  The complex behavior is characteristic of avoided crossings.  The vertical (blue) dashed line is the value of $\log g$ that has the closes period match between the observations and model; this is the model detailed in the third column of Table~\ref{tab01}. } 
\label{fig05} 
\end{figure}

The existence of mixed--character modes in sdB stars leads to avoided crossings and other phenomena associated with mode trapping.  Fig.~\ref{fig05} shows this complex behavior in the shell-burning 0.478M${_\odot}$ model.  This figure clearly shows avoided crossings.  While this complicates the mode identification and analysis, the value of mixed character modes in probing the deep internal structure of sdB stars has enormous potential.  In addition, it provides an opportunity to examine the rotational state of the core to constrain models of angular momentum transport on the HB and during earlier stages (Kawaler \& Hostler 2005).

\section{Conclusions}

Through asteroseismology, we can begin to answer basic questions about stars like our own Sun that have perplexed stellar astrophysics for four decades.  Not only is the origin of the hot subdwarf B stars one of the remaining mysteries of low-mass stellar evolution, but the stars themselves are unique laboratories for studying the physics of diffusion, radiative levitation, hot stellar winds, turbulence, core convection, and semiconvection.  Studies of the cores of these stars can provide a window to these important physical ingredients of a wide variety of stars ranging across the entire H--R diagram.

Many of these stars are in binary systems, and in fact binary evolution may be required to understand their origin.  Since many do reside in close binaries where tidal effects are substantial, we can use the pulsations to probe processes such as interactions between pulsations and tides (Reed et al. 2006), tidal circularization, and spin--orbit synchronization (see for example Charpinet et al. 2008b, van Grootel et al. 2008).

As exciting as the developments in sdB asteroseismology have been, we anticipate that space-based observations with {\it Kepler} will provide revolutionary new data on pulsating sdB stars - in particular, the long-period $g$-mode pulsators.  With these observations in hand, we can look forward to peering at never-before-seen details of the deep interior of stars that represent the future fate of our own Sun.

\acknowledgements
The author is indebted to the organizers of this meeting for their encouragement and assistance in enabling is participation.

\end{document}